\documentclass[10pt,twocolumn]{article}
\usepackage[letterpaper,margin=0.75in,columnsep=0.25in]{geometry}
\usepackage[T1]{fontenc}
\usepackage{times}
\usepackage{microtype}
\usepackage{booktabs}
\usepackage{multirow}
\usepackage{graphicx}
\usepackage{xcolor}
\usepackage{amsmath,amssymb}
\usepackage{listings}
\usepackage{algorithm}
\usepackage{algpseudocode}
\usepackage[numbers,sort&compress]{natbib}
\usepackage[font=small,labelfont=bf,skip=4pt]{caption}
\usepackage{balance}
\usepackage[hidelinks,breaklinks]{hyperref}
\usepackage{url}

\usepackage{titlesec}
\titleformat{\section}{\normalfont\large\bfseries}{\thesection.}{0.5em}{}
\titleformat{\subsection}{\normalfont\normalsize\bfseries}{\thesubsection.}{0.5em}{}
\titleformat{\subsubsection}{\normalfont\normalsize\itshape}{\thesubsubsection.}{0.5em}{}
\titlespacing*{\section}{0pt}{8pt}{4pt}
\titlespacing*{\subsection}{0pt}{6pt}{3pt}

\definecolor{cmtgray}{gray}{0.42}
\definecolor{kwblue}{RGB}{20,60,140}
\definecolor{strgreen}{RGB}{20,110,60}
\definecolor{rulegray}{gray}{0.80}

\lstdefinestyle{py}{
  language=Python,
  basicstyle=\ttfamily\scriptsize,
  keywordstyle=\color{kwblue},
  commentstyle=\color{cmtgray}\itshape,
  stringstyle=\color{strgreen},
  numbers=left,
  numberstyle=\tiny\color{cmtgray},
  numbersep=5pt,
  frame=tb,
  rulecolor=\color{rulegray},
  framesep=3pt,
  columns=fullflexible,
  keepspaces=true,
  showstringspaces=false,
  breaklines=true,
  breakatwhitespace=true,
  captionpos=b,
  aboveskip=6pt,
  belowskip=4pt,
  xleftmargin=10pt,
}
\lstdefinestyle{sh}{
  language=bash,
  basicstyle=\ttfamily\scriptsize,
  keywordstyle=\color{black},
  commentstyle=\color{cmtgray}\itshape,
  stringstyle=\color{black},
  numbers=none, frame=tb, rulecolor=\color{rulegray}, framesep=3pt,
  columns=fullflexible, keepspaces=true, showstringspaces=false,
  breaklines=true, captionpos=b, aboveskip=6pt, belowskip=4pt, xleftmargin=4pt,
}
\newcommand{\code}[1]{\texttt{\small #1}}
\makeatletter
\newcommand{\symfootnote}[2]{\begingroup
  \renewcommand\thefootnote{#1}\footnotetext{#2}\endgroup}
\makeatother

\newcommand{\reptime}{$325$--$332$\,s}
\newcommand{\xstat}{\ensuremath{x}}
\newcommand{\lam}{\ensuremath{\lambda}}

\title{\vspace{-1.5em}\Large\bf Forging Tree-Ring: Reproducing and Instrumenting
Black-Box Semantic Watermark Forgery}

\author{
{\normalsize Saifur Rahman Tamim\textsuperscript{1},
 Md Taslimul Hasan Toufique\textsuperscript{1},
 and A.M.\,Tayeful Islam\textsuperscript{1}}\\[4pt]
{\small Project Group 7\,$^{*}$}\\
{\small \textsuperscript{1}Department of Computer Science and Engineering}\\
{\small Northern University Bangladesh, Dhaka, Bangladesh}\\[3pt]
{\small\ttfamily tamim\_41230201087@nub.ac.bd,\ ripperthemodder@gmail.com}\\
{\small\ttfamily tayef\_nubcse@nub.ac.bd}
}
\date{}

\begin{document}
\maketitle
\symfootnote{$*$}{This paper serves as the final project report for
CSE 4383: Image Processing and Computer Vision Lab Work, Northern University Bangladesh.}

\begin{abstract}
Semantic watermarking schemes such as Tree-Ring hide a detectable pattern in the
initial noise latent of a diffusion model. Recent work shows these watermarks are not
only removable but forgeable: an attacker who never sees the watermarking key can
still produce images the genuine detector accepts. We reproduce the Reprompt forgery
attack of M\"uller et al.\ against Tree-Ring on Stable Diffusion~XL, using the
authors' released code, on free-tier dual T4 GPUs with 14.6\,GB of usable memory per device,
substantially less per-GPU memory than the A40 hardware used in the original study.

The attack reproduces. Over six trials of three arms we detect genuine images $6/6$,
clean images $0/6$, and forged images $5/6$, at \reptime\ per attack. Three further
results came out of running it under constraint. The released detector computes a
non-central $\chi^2$ statistic and hands back only its CDF, so we recovered the
discarded statistic; our recovery matches the released detector exactly, and two
natural scores built from it separate the forged arm from the clean null at AUC
$0.861$ and $0.972$ on the same eighteen observations. Running SDXL in half precision
requires patching the pipeline's direct autoencoder calls, and a controlled probe
confirms the patched path leaves the detector statistic unchanged. Finally, we
report a prediction we made from reading the detector source that our measurements
then contradicted. The notebook, the pinned fork and every measurement artifact are
released with the paper.
\end{abstract}

\section{Introduction}

Text-to-image diffusion models are good enough now that asking whether a picture was
machine-generated has stopped being an academic question. Watermarking is one
prominent answer. Among the watermarking families, semantic schemes are notable
because they add no post-hoc pixel-space perturbation: they encode a signal in the
initial noise latent the image is generated from, so the mark is a property of the
generative trajectory rather than something applied to a finished image. Tree-Ring~\cite{wen2023treering} is
the canonical example, and it survives the compression, cropping and blurring that
defeat pixel-domain marks.

\begin{figure*}[t]
\centering
\includegraphics[width=0.94\textwidth]{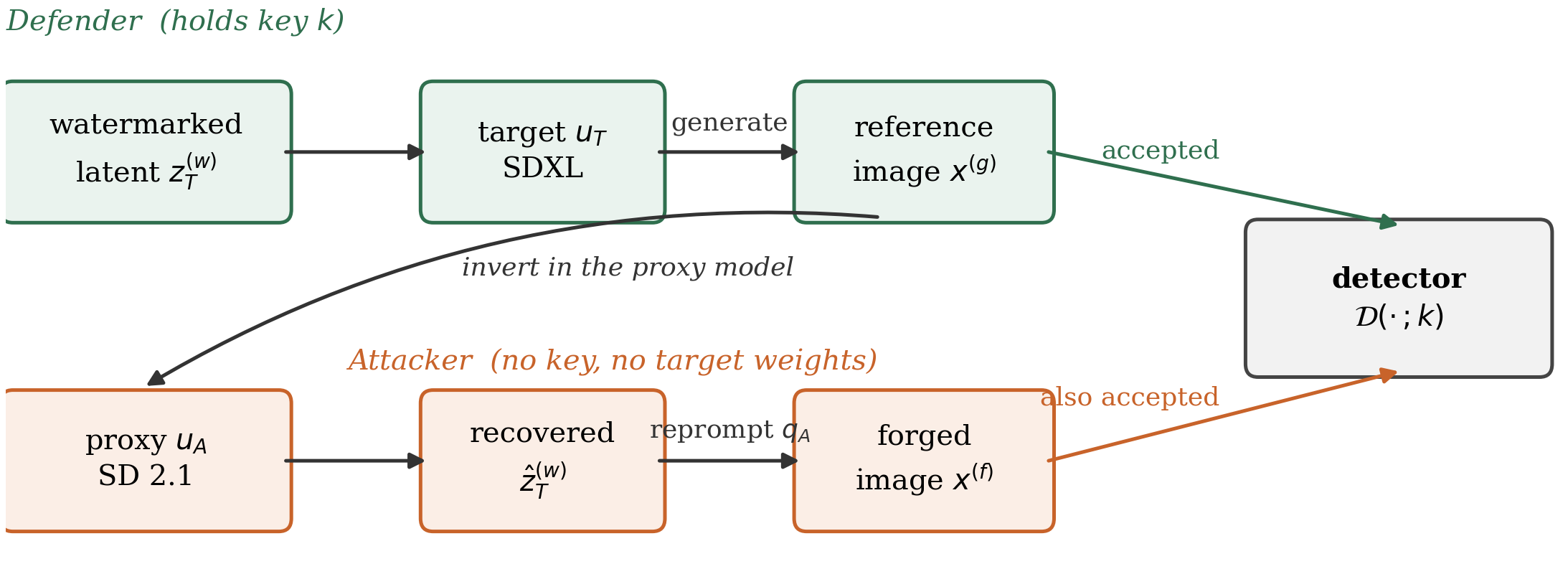}
\caption{The Reprompt forgery attack. The defender generates from a key-derived
latent (top). The attacker takes one reference image, inverts it in a proxy model
that has never seen the key, and regenerates an unrelated image from the recovered
latent (bottom). Both images are accepted by the same detector.}
\label{fig:pipeline}
\end{figure*}

Surviving distortion is not the only thing a provenance mechanism has to do. A
watermark that resists every attack on its robustness is still useless as evidence of
origin if someone can manufacture it. M\"uller et al.~\cite{muller2025semantic} showed
exactly that against Tree-Ring and Gaussian
Shading~\cite{yang2024gaussianshading}. The ring pattern survives DDIM
inversion carried out in a \emph{different} diffusion model, so an attacker who owns
one watermarked reference image can invert it in a proxy model and regenerate
arbitrary new images the genuine detector will accept. No key, no gradients, no
privileged access to the target. Figure~\ref{fig:pipeline} sketches the whole thing.

We set out to reproduce that attack independently. Not because we doubted it, but
because we wanted a verified baseline on hardware we control before building anything
on top of it. Reproduction turns out to be the right first step more often than the
literature admits: a longitudinal study of roughly 750 machine-learning security
papers found that $60\%$ shipped no code, and that $56\%$ of the artifacts which did
exist would not run~\cite{olszewski2023reproducibility}.

Our constraint was compute. M\"uller et al.\ evaluated on NVIDIA~A40 hardware, distributing their
experimental campaign over eight such GPUs; we had a free-tier notebook service
offering two NVIDIA~T4 accelerators with 14.6\,GB of usable memory each. The released code assumes full precision throughout and has no
half-precision path. Most of what we report below came out of working inside that
limit, because it forced us to instrument parts of the pipeline that a bigger machine
would have let us ignore.

\vspace{2pt}
\noindent\textbf{Contributions.}
\begin{itemize}\itemsep2pt
  \item A verified reproduction of Reprompt against Tree-Ring on SDXL under a fully
        pinned software and provenance configuration, with genuine, clean and forged
        detection rates all consistent with the originally reported ones
        (Section~\ref{sec:eval}).
  \item Recovery of the detector's discarded non-central $\chi^2$ statistic, matching
        the released detector exactly, and the observation that two scores derived
        from it give materially different separation on identical samples
        (Section~\ref{sec:instr}).
  \item A mixed-precision compatibility fix for SDXL's direct autoencoder calls,
        with a controlled probe showing the patched path returns the same detector
        output as full precision on the tested sample (Section~\ref{sec:fp16}).
  \item A four-gate feasibility protocol covering memory, cost, statistic
        recoverability and hypothesis separation, reported with measured outcomes and
        including one prediction our data refuted (Sections~\ref{sec:eval}
        and~\ref{sec:disc}).
  \item An exploratory uncertainty analysis of the gate data (bootstrap
        confidence intervals, pairwise rank tests and effect sizes) that bounds
        how much the recovered separations can support at this sample size, and
        finds that bound to be low (Section~\ref{sec:eval}).
\end{itemize}

\section{Background and Related Work}

\subsection{Semantic watermarking}

Frequency-domain watermarking predates diffusion models by decades~\cite{cox2007digital},
and the learned pixel-space variants that followed, such as
HiDDeN~\cite{zhu2018hidden}, embed a message with a jointly trained
encoder-decoder pair. A third family fine-tunes the generator itself so that its
outputs carry a signature, as in the Stable
Signature~\cite{fernandez2023stablesignature}. Semantic schemes take a fourth route.

A latent diffusion model~\cite{rombach2022ldm,ho2020ddpm} builds an image by
repeatedly denoising an initial Gaussian latent $z_T$. Tree-Ring~\cite{wen2023treering} watermarks the
model by overwriting part of $z_T$ in the Fourier domain with a fixed, key-derived
pattern: concentric rings placed in one channel of the shifted spectrum. Since the
pattern goes in before generation rather than after, it is carried through the entire
sampling trajectory and is not localised anywhere in the output.

Detection runs the process backwards. The detector inverts the sampler with
DDIM~\cite{song2021ddim} to estimate the latent the image would have come from,
applies the same frequency-domain mask, and compares the recovered coefficients
against the key pattern. With $f$ the recovered masked coefficients, $t$ the key
pattern and $\sigma$ the standard deviation of $f$, the released implementation
computes
\begin{equation}
\xstat = \sum_i \left( \frac{f_i - t_i}{\sigma} \right)^{\!2},
\qquad
\lam = \sum_i \frac{t_i^{2}}{\sigma^{2}},
\label{eq:trstat}
\end{equation}
and reports $p = F_{\chi'^2}(\xstat;\, \mathrm{df},\, \lam)$, where $\mathrm{df}$ counts the
real and imaginary masked coefficients. A detection is declared
when $p < \tau$; for SDXL~\cite{podell2024sdxl} at a $1\%$ false-positive rate,
$\tau = 0.0261$. One detail matters later: detection works on an \emph{estimate} of
$z_T$, since DDIM inversion is only approximately invertible. Every detection is
already one inference removed from the quantity the scheme embeds.

Ci et al.~\cite{ci2024ringid} revisited the scheme and found that part of its
robustness comes not from pattern matching but from a distribution shift the
watermarking process introduces incidentally. They also showed that Tree-Ring
handles multiple distinct keys poorly, where that shift gives no help, and proposed
RingID to address it. Both observations bear on our own single-key limitation
(Section~\ref{sec:limits}).

\subsection{Removal versus forgery}

Most of the attack literature goes after removal, destroying the mark while keeping
the image usable. Zhao et al.~\cite{zhao2024removable} give the strongest general
result in that direction: adding noise in latent space and regenerating provably
removes pixel-level watermarks, though they observe that semantically-embedded marks
resist the same attack, which is part of why semantic schemes drew attention in the
first place. The WAVES benchmark~\cite{an2024waves} organised this into a
stress-test suite spanning distortion, regeneration and adversarial attacks, and found
Tree-Ring especially vulnerable to adversarial embedding and surrogate-detector
attacks while holding up reasonably well against plain distortion.

Forgery runs the other way, and for provenance it is the more damaging direction.
Removal causes false negatives: a genuine image stops being recognised. Forgery causes
false positives: an image the marked model never produced gets attributed to it. A
provenance claim depends entirely on the second error rate, and robustness results say
nothing about it.

\subsection{The Reprompt attack}

M\"uller et al.~\cite{muller2025semantic} demonstrate that the ring pattern survives
DDIM inversion in a model other than the one that generated the image. The attacker
holds a watermarked reference $x^{(w)}$, inverts it in a proxy $u_A$ to get
$\hat{z}_T^{(w)}$, then generates a new image from that latent under whatever prompt
they like. The output looks nothing like the reference and the genuine detector
accepts it anyway.

The same paper introduces Imprint, an optimisation-based variant that learns a
perturbation to a cover image's latent so its inversion matches a target watermarked
latent. Imprint buys finer control at roughly two orders of magnitude more cost per
run. We reproduce Reprompt only, for reasons given in Section~\ref{sec:limits}.

Later work has chased cheaper forgery. Zhu et al.~\cite{zhu2025optfree} propose
Plug-and-Plant, an optimisation-free attack that extracts and re-implants a watermark
by regenerating the image through an off-the-shelf diffusion model, reporting up to
$100\%$ detectability across two dozen model-data-watermark combinations. Zhou et
al.~\cite{zhou2026metaseal} approach the problem from the defensive side with
MetaSeal, arguing that content-agnostic marks verified by a detector are structurally
forgeable and that cryptographic, content-dependent signatures are the way out. Their
framing is useful to us: it identifies the two properties that make a scheme
forgeable in the way Tree-Ring is. Zhang et al.~\cite{zhang2026sembind} attack the same problem
from the generative side with SemBind, binding the latent watermark signal to image
semantics through a contrastively trained masker so that a latent lifted from one
image no longer verifies against unrelated content. Closest to our own observations,
Lee et al.~\cite{lee2026geometric} model black-box forgery as a rate-distortion problem and
argue that proxy-target model mismatch imposes an irreducible distortion floor on how
faithfully a watermark can be forged, manifesting as structured geometric deviation
on the latent manifold rather than as noise, which they exploit for scheme-agnostic
detection ahead of watermark verification. Our forged arm occupying a measurable band
between the clean null and the genuine arm (Section~\ref{sec:eval}) is an empirical
signature consistent with such a floor, arrived at independently and by measurement alone.

\subsection{Reproduction as a research activity}

Independent reproduction is an established genre. Olszewski et
al.~\cite{olszewski2023reproducibility} measured its state across a decade of Tier-1
security venues, finding that $60\%$ of papers shipped no code and that $56\%$ of the
artifacts which did exist would not run. They further report no statistically
significant change in artifact \emph{availability} after artifact evaluation
committees were introduced, while noting that artifacts which passed evaluation
worked at a higher rate than those that did not.

Reproductions that diverge are informative in their own right. Hwang et
al.~\cite{hwang2024invisible}, reproducing Tree-Ring on the removal side, matched the
original's unattacked performance closely but got an AUC of $0.463$ under rotation
where the original reported $0.935$. Rather than bury it, they traced the gap to
rotation in pixel space not corresponding to rotation in latent space, and confirmed
the effect held across channel placements, model versions and schedulers. We take the
same posture below: where our measurements diverge, we say so.
\section{Reproduction Setup}
\label{sec:setup}

\subsection{Hardware}

\begin{table}[t]
\centering\small
\caption{Original and reproduction environments.}
\label{tab:hw}
\begin{tabular}{lll}
\toprule
 & M\"uller et al. & This work \\
\midrule
GPU         & A40, 48\,GB nominal & T4 $\times 2$, 14.6\,GB ea. \\
Precision   & fp32            & fp16 target / fp32 attacker \\
Target      & SDXL            & SDXL \\
Attacker    & SD 2.1          & SD 2.1 \\
Resolution  & 512             & 512 \\
Scheduler   & DDIM            & DDIM \\
Allocation  & dedicated       & 30 GPU-h / week, shared \\
\bottomrule
\end{tabular}
\end{table}

Table~\ref{tab:hw} sets out the gap we were working across. The notebook service
offers two accelerator types, and one of them is unusable: the P100 reports compute
capability \code{sm\_60}, which the installed PyTorch~2.10 build has dropped support
for. That leaves the dual-T4 configuration. Neither T4 supports \code{bfloat16},
which puts FLUX.1-dev out of reach even though the released code supports it. This is
a hardware exclusion and says nothing about how the scheme behaves on that model.

Storage is tiered, and the tiers have different lifetimes. Model weights go to an
ephemeral overlay volume of roughly 1\,TB; experimental outputs go to a 20\,GB
persistent volume. Section~\ref{sec:env} describes a failure we caused by treating
those two as interchangeable.

\subsection{Models and configuration}

The target is Stable Diffusion~XL base~1.0 on a DDIM scheduler and the attacker proxy
is Stable Diffusion~2.1 base. Stability withdrew the original SD~2.1 repository while
we were working, so we use a mirror of the same weights. Generation and inversion both
run at $512 \times 512$ with $50$ steps and guidance scale $7.5$. Tree-Ring uses the
released defaults: ring pattern, circular mask, radius $10$, channel $3$, complex
injection, \code{l1\_complex} measurement, giving $\mathrm{df} = 634$. The SDXL
threshold is $\tau = 0.0261008646426459$, taken from the original calibration at a
$1\%$ false-positive rate.

Two seeds are in play and they should not be confused. The generation seed is $123 + i$ for
trial $i$, so the six trials use $123$ through $128$. The Tree-Ring
key is \emph{not} varied: \code{w\_seed} is fixed at $999999$ for every trial. All
eighteen observations therefore share one watermark key, which is the study's
sharpest limitation and is discussed in Section~\ref{sec:limits}.

The recorded session ran PyTorch~2.10.0+cu128 on CUDA~12.8, Python~3.12.13,
\code{transformers}~4.46.3 and \code{diffusers}~0.31.0.

\subsection{Running the released attack}

For reference, and because it is the shortest possible statement of what the original
artifact does, the attack is invoked as a single command:

\begin{lstlisting}[style=sh,caption={The released Reprompt entry point, exactly
as our notebook invokes it. Everything else in this paper is reproduction and
instrumentation work built around this one command.},label=lst:cli]
python run_reprompting.py \
  --wm_type TR \
  --modelid_target \
    stabilityai/stable-diffusion-xl-base-1.0 \
  --modelid_attacker \
    Manojb/stable-diffusion-2-1-base \
  --resolution 512 --seed 123
\end{lstlisting}

\subsection{Provenance}

All our modifications live in a fork of the released
repository~\cite{semanticforgery-repo}, pinned by commit hash and listed in the
Availability section. We chose that over
runtime patching for two reasons. Monkey-patching in the notebook process is invisible
to anything launched as a subprocess, so a patched detector in the parent does not
apply to the attack script it starts. And a dirty working tree satisfies a commit-hash
check while running code that is not in the recorded commit. The notebook therefore
issues \code{git reset -{}-hard} to the pinned SHA followed by \code{git clean -fd},
re-reads \code{HEAD}, and asserts both that it matches and that
\code{git status -{}-porcelain} comes back empty.

Session metadata is serialised at the start of every run: commit hashes for our fork
and the upstream base, model identifiers, dtypes, device assignment, library versions,
and the SHA-256 of the resolved dependency lock. It ships with the artifacts.

\subsection{Gates}

We ran the reproduction as a sequence of feasibility gates rather than as an
experiment. Each one asks a question that would make the rest pointless if answered
badly, and none is meant to produce a finding about watermarking:

\begin{itemize}\itemsep2pt
\item[\textbf{G1}] Do both pipelines fit on the hardware?
\item[\textbf{G2}] What does one attack cost in wall-clock time?
\item[\textbf{G3}] Can the detector's underlying statistic be recovered, and does it
      carry usable range?
\item[\textbf{G4}] Do genuine, clean and forged observations separate?
\end{itemize}

\section{Environment Reconstruction}
\label{sec:env}

Four things broke before we had a usable pipeline. We report them as findings rather than as a
narrative because each one generalises past this particular repository.

\subsection{Package removal is not idempotent in layered images}

The notebook image ships PyTorch~2.10 and \code{transformers}~v5; the released code
wants the late-2024 stack. Installing the repository's pinned requirements on top
produces a tree where v5 and v4 files sit side by side, because \code{pip} cannot
fully remove distributions baked into a lower image layer. Every entry point then dies
at import with a \code{ModuleNotFoundError} naming a module the codebase never
references, which sends you looking for a missing dependency instead of a corrupted
one.

Getting out requires uninstalling, deleting the package directories from
\code{site-packages} by hand, installing five exact pins
(\code{transformers}~4.46.3, \code{tokenizers}~0.20.3, \code{diffusers}~0.31.0,
\code{huggingface\_hub}~0.26.2, \code{accelerate}~1.1.1), and restarting the
interpreter. The restart is not optional, since the half-imported broken modules stay
cached in \code{sys.modules} and survive reinstallation.

One dependency is easy to overlook. \code{lpips} is in the requirements file but not
among the version-critical pins, and without it every entry point fails inside the
image-utility module before a model is ever loaded.

\subsection{Cache paths are frozen at import}

The Hugging Face hub library resolves its cache location the first time it is
imported. Setting \code{HF\_HOME} after any transitive import does nothing, so weights
land in the default location, which for us was the 20\,GB persistent volume they then
filled. The variable has to be set in the first executed cell, above every import in
the session. We point weights at the ephemeral high-capacity volume and keep the
persistent one for outputs.

\subsection{Guards must interrogate state, not sentinel files}

Our first version of the pinning cell wrote a lock file on success and skipped
reinstallation whenever that file existed. The two objects have different lifetimes:
the persistent volume survives across sessions, installed packages do not. So a stale
lock file caused the pinning step to be skipped in an environment that no longer had
the pins, and the session carried on with the platform's stock \code{diffusers}. The
guard reported success the whole time. We only caught it by reading library versions
out of the session metadata.

The fixed guard calls \code{importlib.metadata.version} on each pinned distribution
and compares against the requirement, treating \code{PackageNotFoundError} as failure.
The general lesson is that a guard has to interrogate the state it claims to protect.
A marker stored on a medium with a different lifetime than the guarded state will
eventually tell you something false.

\subsection{Half precision needs the autoencoder handled explicitly}
\label{sec:fp16}

The released code sets \code{DTYPE = torch.float32} at module level in its pipeline
provider and has no half-precision path at all. We introduced fp16 on the target to
buy headroom. Detection then returned \code{NaN} latents while the image-quality
metrics stayed well-formed, which told us generation was fine and only the detection
path was broken.

The cause is specific. SDXL's autoencoder carries \code{force\_upcast = True} in its
configuration, and the \code{diffusers} pipeline honours that flag inside its own
\code{\_\_call\_\_}. The repository, however, calls the autoencoder directly from
\code{vae\_encode} and \code{latents\_to\_imgs}, which bypasses that handling
entirely. In fp16 those direct calls overflow. Our fork adds a
\code{\_vae\_upcast} helper that saves the autoencoder's dtype, promotes it to fp32
when it is fp16 and \code{force\_upcast} is set, and restores it afterwards, wrapped
around every direct encode and decode.

To check the patched path we built a probe that splits inversion into its two stages,
\code{imgs\_to\_latents} (autoencoder encode) and \code{invert\_z0} (the U-Net DDIM
loop), and runs each under both dtypes against one fixed generated image so the
autoencoder dtype is the only thing changing. After the patch, neither stage produced
non-finite values under either configuration and both returned the same detection
$p$-value of $1.6648 \times 10^{-36}$ to the last digit.

That result validates the fix. It does not show that an unpatched fp16 autoencoder
would have been safe, and we want to be careful about the difference: the probe ran
against a fork that already contains the upcast, so what it demonstrates is that the
mixed-precision path preserves detector output on the tested image, not that the
original overflow was imaginary. The attacker model stays in fp32 regardless. It needs
the headroom through a $50$-step inversion chain and at $5.57$\,GB there is nothing to
gain from shrinking it.
\section{Detector Instrumentation}
\label{sec:instr}

\subsection{The detector discards its own statistic}

The released Tree-Ring detector computes $\sigma$, $\lam$ and $\xstat$ exactly as in
Equation~\ref{eq:trstat}, evaluates the non-central $\chi^2$ CDF, and returns only the
resulting $p$-value; its accuracy accessor returns $1 - p$. The intermediate
quantities are computed and thrown away.

For a detector whose job is to accept or reject, that is a sensible interface. It gets
limiting as soon as a consumer needs to compare evidence strengths instead of
outcomes. A CDF value is a tail probability, so the whole range of strong evidence is
compressed into a narrow band next to zero; the accuracy accessor saturates at $1.0$
and loses more still. More simply, $\xstat$, $\lam$, $\mathrm{df}$ and $\sigma$ are all
computed and then not returned, and differences in $p$ do not correspond to comparable
differences in how well the recovered spectrum actually matches the key. We had also
expected the compression to produce outright numerical censoring; it did not, and
Section~\ref{sec:g3} reports that measurement.

So we reimplemented Equation~\ref{eq:trstat} with $\xstat$, $\lam$, $\mathrm{df}$ and
$\sigma$ all kept (Listing~\ref{lst:rawstat}). The computation is the detector's; only
the return value differs.

\begin{lstlisting}[caption={Recovering the discarded statistic. The arithmetic mirrors
the released detector; the intermediates are kept rather than collapsed into a
CDF.},label=lst:rawstat]
def raw_tr_stat(wm, latents):
    lat_fft = torch.fft.fftshift(
        torch.fft.fft2(latents), dim=(-1, -2))
    mask, gt = wm.watermarking_mask[0], wm.gt_patch[0]
    out = []
    for one in lat_fft:
        f = one[mask].flatten(); t = gt[mask].flatten()
        f = torch.concatenate([f.real, f.imag])
        t = torch.concatenate([t.real, t.imag])
        sigma = f.std()
        lambd = (t ** 2 / sigma ** 2).sum().item()
        x     = (((f - t) / sigma) ** 2).sum().item()
        out.append({"x": x, "lambd": lambd, "df": len(t),
                    "sigma": sigma.item(),
                    "p_value": scipy.stats.ncx2.cdf(
                        x=x, df=len(t), nc=lambd)})
    return out
\end{lstlisting}

\subsection{Validating the reimplementation}

A reimplemented statistic is only useful if it agrees with the one the detector
applies. On every observation we push our recovered $\xstat$ back through the same CDF
and compare against what the released detector returns for that latent, asserting
agreement below $10^{-12}$. The assertion held on all eighteen observations at a
maximum absolute difference of exactly $0$. The arithmetic is identical rather than
approximate, which rules out reimplementation error as an explanation for anything in
Section~\ref{sec:eval}.

\subsection{The choice of score is a modelling decision}
\label{sec:score}

Two scores fall out of Equation~\ref{eq:trstat} naturally. The raw discrepancy
$-\xstat$ is the simpler one: lower $\xstat$ means a closer match to the key. The
adjusted form $\lam - \xstat$ offsets the discrepancy by the non-centrality parameter,
which is itself image-dependent through $\sigma$ and so varies between observations.

\begin{figure}[t]
\centering
\includegraphics[width=\columnwidth]{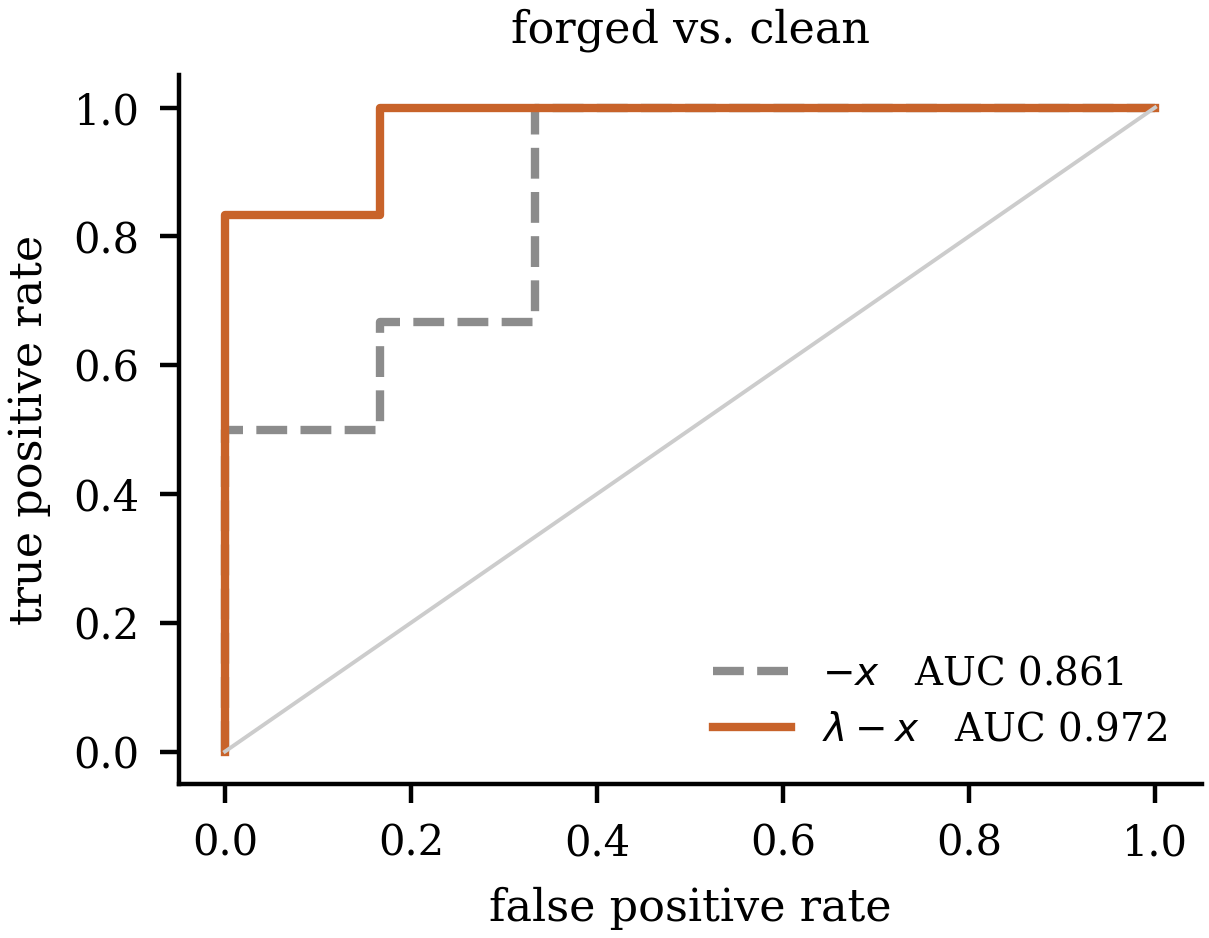}
\caption{Separating forged from clean under the two candidate scores. Both are derived
from the same eighteen observations; only the transformation differs.}
\label{fig:roc}
\end{figure}

They are not the same thing written two ways. Across our runs $\sigma$ spans $31.75$
to $42.59$ and $\lam$ spans $856.6$ to $1541.4$, so the two scores are not
monotonically related. On the same eighteen observations the forged arm separates from
the clean null at AUC $0.861$ under $-\xstat$ and $0.972$ under $\lam - \xstat$; the
genuine arm reaches $1.000$ under both. Figure~\ref{fig:roc} shows the two curves.

A gap that size, produced purely by a choice of transformation, means the score has to
be fixed in advance rather than chosen once separation is visible. Picking between
candidate scores after seeing how each performs on the samples you then report is
selection on the test set. We report both and treat neither as canonical.

\subsection{Measurement harness}

The released attack script writes only the $p$-value to its metrics output,
so our measurement runs do not go through it. We assemble the attack from the
repository's own primitives, with inversion, generation and detection all unmodified,
and add a third arm the script does not produce. Each trial yields three observations
against a common watermark instance:

\begin{itemize}\itemsep2pt
\item \textbf{genuine}: a watermarked latent generated in the target and inverted
      back;
\item \textbf{clean}: the same prompt generated from ordinary Gaussian noise, an
      empirical null for both other arms;
\item \textbf{forged}: the genuine image inverted in the attacker model,
      regenerated there under an unrelated prompt, and handed back to the target
      detector.
\end{itemize}

The clean arm is the addition that earns its place. Without an empirical null there is
no way to say what a given score is evidence \emph{for}; you can only say whether it
crossed a threshold.

\begin{algorithm}[t]
\small
\caption{One measurement trial. $u_T$ is the target, $u_A$ the attacker proxy,
$\mathcal{I}$ inversion, $\mathcal{G}$ generation, $\mathcal{D}$ the Tree-Ring
detector and $k$ the key. Prompts $q_T$ and $q_A$ are unrelated.}
\label{alg:trial}
\begin{algorithmic}[1]
\State $z_T^{(w)} \gets \textsc{Watermark}(k)$ \Comment{key-derived latent}
\State $x^{(g)} \gets \mathcal{G}(u_T, q_T, z_T^{(w)})$ \Comment{genuine}
\State $x^{(c)} \gets \mathcal{G}(u_T, q_T, \varepsilon),\;
        \varepsilon \sim \mathcal{N}(0, I)$ \Comment{clean null}
\State $\hat{z}^{(w)} \gets \mathcal{I}(u_A, x^{(g)})$
        \Comment{invert in \emph{attacker} model}
\State $x^{(f)} \gets \mathcal{G}(u_A, q_A, \hat{z}^{(w)})$ \Comment{forged}
\For{$x \in \{x^{(g)}, x^{(c)}, x^{(f)}\}$}
  \State $\hat{z} \gets \mathcal{I}(u_T, x)$
  \State record $(\xstat, \lam, \mathrm{df}, \sigma, p)
         \gets \mathcal{D}(\hat{z}, k)$
\EndFor
\end{algorithmic}
\end{algorithm}

Algorithm~\ref{alg:trial} states one trial. Line~4 is the whole vulnerability: the
ring pattern survives inversion through a model that never held the key, so line~5 can
regenerate an unrelated image from a latent that still carries it. The two models sit
on separate devices and are swapped to host memory around each other's turn.

To be explicit about the boundary: the attack concept, the DDIM inversion, the
Tree-Ring detector, the diffusion models and the prompt sets are M\"uller et al.'s.
The loop assembly, the clean null arm, the recovery of the discarded statistic and its
validation are ours.
\section{Evaluation}
\label{sec:eval}

The full pipeline was executed twice in independent sessions eleven days apart, from a
clean environment build each time. Every statistic reported below reproduced to the last
significant digit across both runs; wall-clock time was the only quantity that varied, by
$2.2\%$. Figures quoted here are from the first session.

All results below come from six trials of three arms, $n = 18$ observations, at
generation seeds $123$ through $128$ under the single fixed Tree-Ring key.

\subsection{G1: memory feasibility}

\begin{figure}[t]
\centering
\includegraphics[width=\columnwidth]{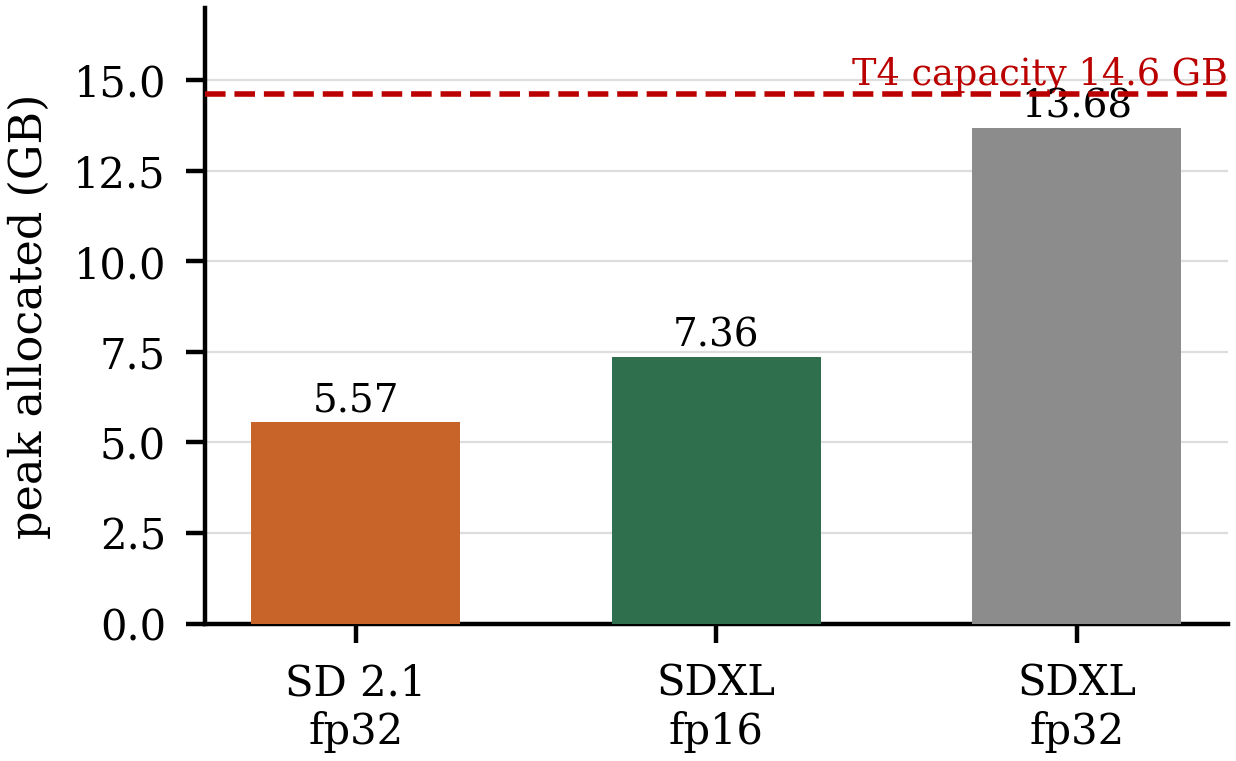}
\caption{Peak allocated memory per pipeline, from
\code{torch.cuda.max\_memory\_allocated} after one generation. Every configuration
fits on a single T4 individually; half precision on the target frees $6.32$\,GB.}
\label{fig:mem}
\end{figure}

Figure~\ref{fig:mem} gives the measured peaks. The attacker at fp32 reaches
$5.57$\,GB, the target at fp16 reaches $7.36$\,GB, and the target at fp32 reaches
$13.68$\,GB. All three fit on a single 14.6\,GB card individually, which is what
matters here: our configuration places the attacker on \code{cuda:0} and the target on
\code{cuda:1}, so the relevant question is per-device headroom rather than the sum.
\textbf{G1 passes.}

Half precision on the target is therefore not required for Reprompt to run on this
hardware. What it buys is headroom: $6.32$\,GB of it, leaving roughly $7.2$\,GB free
on the target device rather than under $1$\,GB. We adopted it for that margin, not
out of necessity.

\subsection{G2: wall-clock cost}

One Reprompt run through the released entry point completes in \reptime\ on a T4, measured across two independent sessions ($332.0$\,s and $324.8$\,s).
M\"uller et al.\ report roughly $30$ seconds for the smaller models and up to two
minutes for FLUX.1 in single-GPU, single-batch runs, with their full experimental
campaign spread over eight A40s. Our figure is therefore several times their published
single-GPU range, which is in line with the gap in memory bandwidth and tensor-core
generation between the two cards. \textbf{G2 passes.}

The number that matters for a threat model is not the ratio but the absolute cost.
Forgery is a sub-six-minute operation on hardware anyone can access for free.

\subsection{G3: statistic recoverability and range}
\label{sec:g3}

All eighteen observations produced distinct values of $\xstat$, spanning $889.1$ to
$1802.2$ against $\mathrm{df} = 634$, and the cross-check against the released
detector returned a maximum absolute difference of exactly $0$. The statistic is
recoverable and carries range. \textbf{G3 passes.}

The $p$-value did not behave the way we predicted. From reading the detector source we
had inferred that \code{scipy.stats.ncx2.cdf} would underflow to exactly $0.0$ on
strong genuine detections, left-censoring the genuine distribution and making
$-\log_{10} p$ undefined, and we had used that inference to justify recovering
$\xstat$ in the first place. What we measured: $p$ was exactly zero in $0$ of $18$
observations, all eighteen values were distinct, and the smallest was
$5.45 \times 10^{-49}$. The prediction failed. The case for recovering the raw
statistic still stands on the grounds in Section~\ref{sec:score}, but not on the
grounds we originally gave for it.

\subsection{G4: hypothesis separation}

\begin{table}[t]
\centering\small
\caption{Per-arm outcomes ($n = 6$ each) beside the values M\"uller et al.\ report for
the same model pairing.}
\label{tab:arms}
\begin{tabular}{lrrrr}
\toprule
Arm & $\bar{\xstat}$ & $\overline{\lam - \xstat}$ & detected & orig.\ rate \\
\midrule
genuine & $1096.5$ & $+220.0$ & $6/6$ & $1.00$ \\
clean   & $1623.9$ & $-626.3$ & $0/6$ & $0.01$ \\
forged  & $1391.2$ & $-371.3$ & $5/6$ & $0.97$ \\
\bottomrule
\end{tabular}
\end{table}

\begin{table}[t]
\centering\small
\caption{Detection $p$-values. Means are dominated by the largest observation in each
arm, so medians are given alongside.}
\label{tab:pvals}
\begin{tabular}{lrrr}
\toprule
Arm & median $p$ & mean $p$ & orig.\ mean $p$ \\
\midrule
genuine & $8.63 \times 10^{-38}$ & $7.96 \times 10^{-24}$ & $1.10 \times 10^{-21}$ \\
clean   & $4.56 \times 10^{-1}$  & $4.80 \times 10^{-1}$  & $4.72 \times 10^{-1}$ \\
forged  & $2.30 \times 10^{-5}$  & $1.50 \times 10^{-2}$  & $5.20 \times 10^{-3}$ \\
\bottomrule
\end{tabular}
\end{table}

\begin{figure}[t]
\centering
\includegraphics[width=\columnwidth]{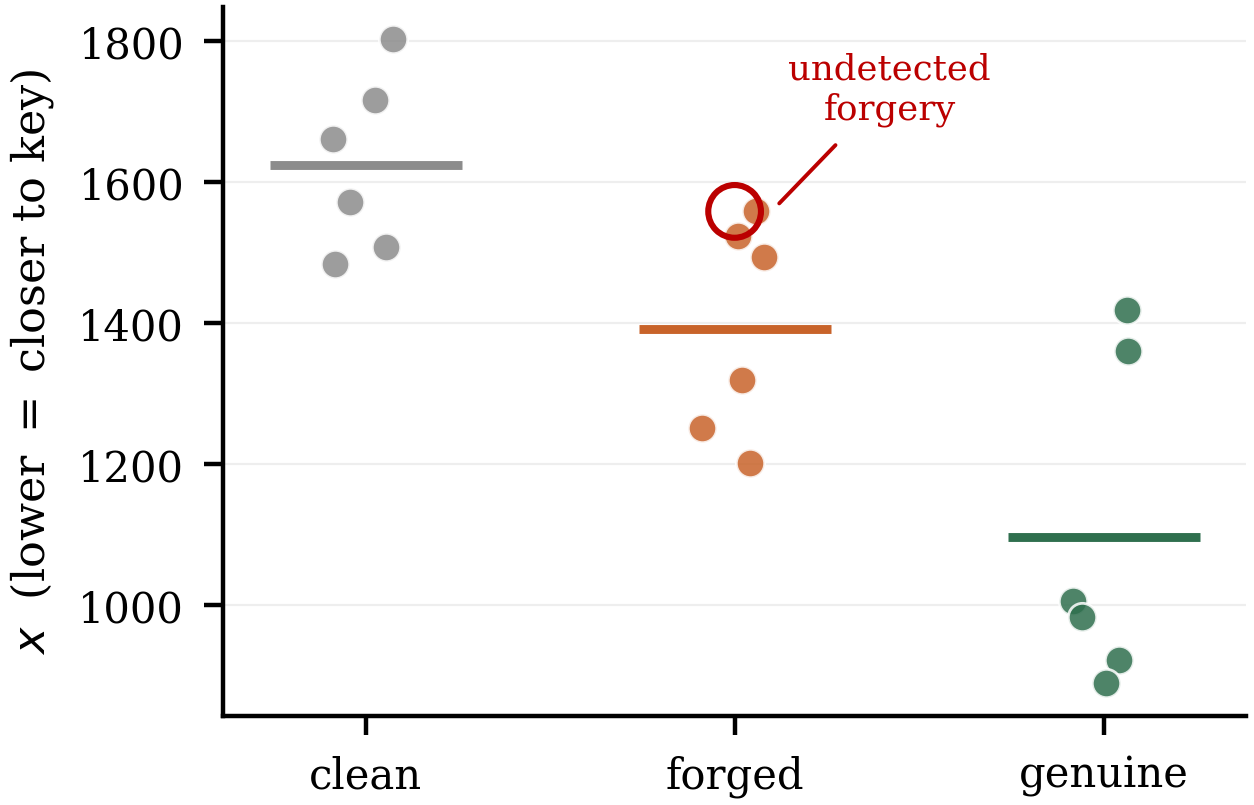}
\caption{Recovered statistic $\xstat$ by arm, one point per trial, with arm means
marked. The forged arm sits between the clean null and the genuine arm. The circled
point is the single forgery the detector missed; its statistic falls inside the clean
range.}
\label{fig:stat}
\end{figure}

\begin{figure}[t]
\centering
\includegraphics[width=\columnwidth]{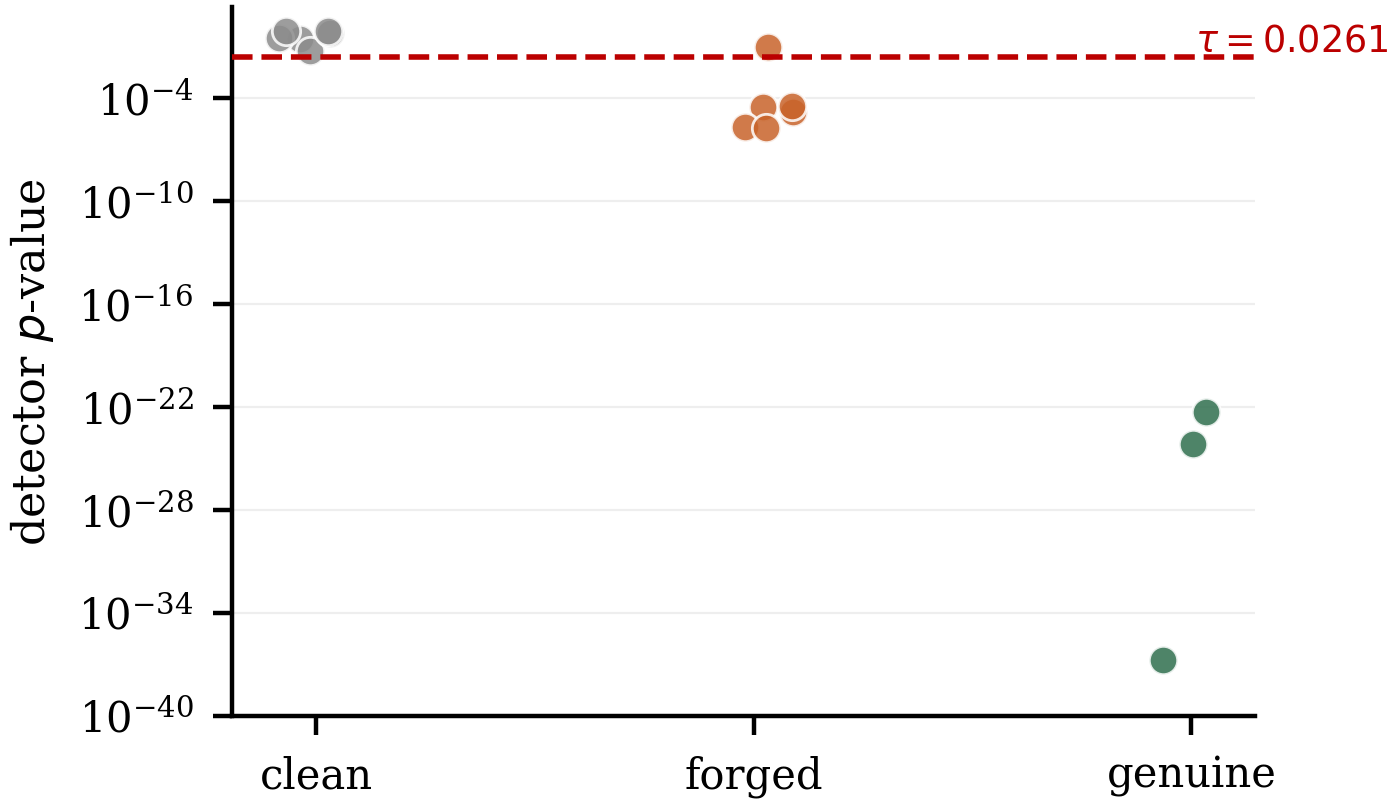}
\caption{Detector $p$-values on a log scale against the calibrated threshold
$\tau = 0.0261$. No observation reached exactly zero; the smallest was
$5.45 \times 10^{-49}$.}
\label{fig:pvals}
\end{figure}

\paragraph{The attack reproduces.} Tables~\ref{tab:arms} and~\ref{tab:pvals} put our
measurements next to the original work's. The clean arm lines up closely, with a mean
$p$ of $0.480$ against $0.472$ and no false positives in six trials against a reported
rate of $0.01$. That is the arm that most directly validates the detector as we
configured it. The genuine arm is detected every time. The forged arm crosses
$\tau = 0.0261$ in five of six trials at a median $p$ of $2.30 \times 10^{-5}$, three
orders of magnitude clear of the threshold (Figure~\ref{fig:pvals}). Running the released entry point directly
gave the same picture: a genuine $p$ of $3.53 \times 10^{-33}$ and a forged $p$ of
$3.13 \times 10^{-6}$, both detected.

\begin{figure*}[t]
\centering
\includegraphics[width=0.94\textwidth]{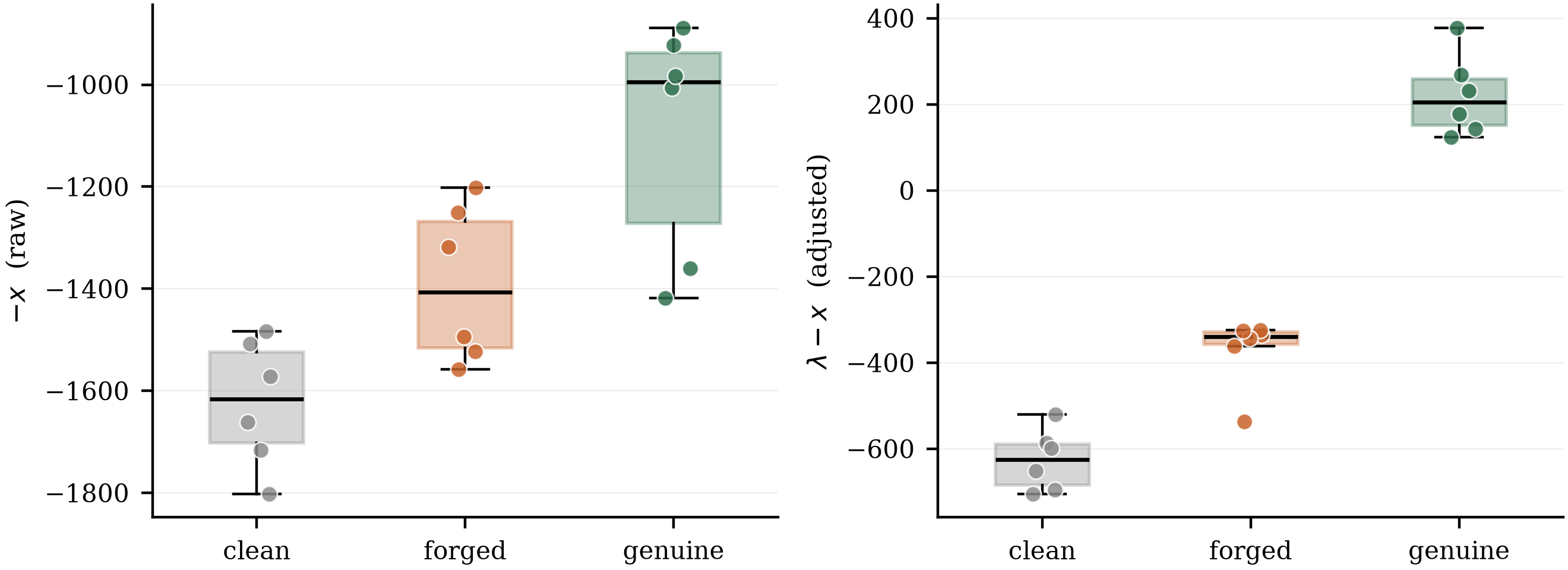}
\caption{Per-arm distributions of both candidate scores, boxes spanning the
interquartile range with the individual trials overlaid. The genuine arm clears
the clean null on both scores; the forged arm occupies the gap between them.}
\label{fig:dist}
\end{figure*}

\paragraph{Separation.} Against the clean null the genuine arm separates perfectly,
AUC $1.000$ on both scores. The forged arm reaches $0.861$ on $-\xstat$ and $0.972$ on
$\lam - \xstat$. \textbf{G4 passes.}

\paragraph{Exploratory uncertainty analysis.} Because $n = 6$ per arm can make a
single AUC look more conclusive than it is, we also ran a small battery of post-hoc
tests over the same eighteen observations, all of which are reproduced by the accompanying
\code{stat\_analysis.py} on the published \code{phase0\_scores.csv}. A $95\%$
bootstrap interval on the forged-vs-clean AUC spans
$[0.58, 1.00]$ for $-\xstat$ and $[0.83, 1.00]$ for $\lam - \xstat$. A two-sided
Mann-Whitney test separates the forged arm from the clean null at $p = 0.041$
and the genuine arm from the clean null at $p = 0.002$, with the forged and genuine arms sitting
$1.67$ and $2.84$ pooled standard deviations below the null in the raw statistic
(Cohen's $d$). The genuine-forged comparison is the loosest of the three at
$p = 0.065$. That ordering is Figure~\ref{fig:stat}'s intermediate band in
different clothes: the forged arm is measurably closer to valid evidence than
the null is, while still not being confusable with a genuine image on six draws.
Figure~\ref{fig:dist} shows both candidate scores side by side.

These values bound the evidence rather than establish it, and we report them as
exploratory throughout. Three limits apply at once. The observations are matched by
trial (the forged image of trial $i$ is derived from the genuine image of that
trial, and the clean image shares its prompt), so an unpaired test is not the
natural one for this design. The sample size puts both candidate tests near their
resolution limits: at $n = 6$ per arm the smallest attainable two-sided Mann-Whitney
$p$ is $2/\binom{12}{6} = 0.0022$ and the smallest attainable exact paired Wilcoxon
$p$ is $2/2^{6} = 0.0312$, so a value at either floor records complete separation and
nothing finer. And the three comparisons are uncorrected; under a Bonferroni
correction at $\alpha = 0.05/3$ none of the paired tests survives on either score,
which the $0.0312$ floor already guarantees, while the unpaired tests survive on
$\lam - \xstat$ but not on $-\xstat$.

Which conclusion one reaches therefore depends on the score and on the test, and we
are not in a position to adjudicate either at eighteen observations. We draw no
inferential conclusion from these numbers and rely on the bootstrap intervals and
descriptive effect sizes instead, which is also why the AUC values in
Section~\ref{sec:score} carry the argument about score choice rather than these
tests.

\paragraph{One divergence worth naming.} Our forged detection rate of $5/6 = 0.83$
comes in under the $0.97$ reported originally. At $n = 6$ that difference is not
meaningful; a $95\%$ Wilson score interval~\cite{wilson1927interval} on $5/6$ runs from $0.44$ to $0.97$ and contains
their value. We are not claiming the attack is weaker than reported.

The shape of the failure is more interesting than the rate. Figure~\ref{fig:stat}
shows the forged arm occupying a band between the clean null and the genuine arm, and
the one forgery the detector missed has $\xstat = 1558.3$, sitting inside the clean
range of $1483.7$ to $1802.2$. It did not fail narrowly; on the statistic the detector
actually uses, that image was indistinguishable from an unwatermarked one.

That intermediate band is the observation we would carry forward. A detector working
as a binary test reports the five successful forgeries and the six genuine images
identically. Whatever would distinguish them is visible in the raw statistic and gone
from the thresholded decision, and the detector computes it either way.
\section{Discussion}
\label{sec:disc}

\subsection{Where the effort actually went}

Invoking the released attack takes one command, shown in Listing~\ref{lst:cli}. Our
reproduction notebook is $375$ lines. Table~\ref{tab:effort} breaks that down, and the
ratio is the part worth saying out loud: the published artifact is not the expensive
part of reproducing a published artifact.

\begin{table}[t]
\centering\small
\caption{Distribution of reproduction code by purpose.}
\label{tab:effort}
\begin{tabular}{lrl}
\toprule
Component & Lines & Category \\
\midrule
Environment repair, provenance & 132 & portability \\
Memory measurement (G1)        & 36  & instrumentation \\
Invoking the released attack   & 1   & original artifact \\
Timing wrapper (G2)            & 22  & instrumentation \\
Precision patch and probe      & 49  & portability \\
Statistic recovery and harness & 71  & extension \\
Analysis and gate verdicts     & 64  & extension \\
\bottomrule
\end{tabular}
\end{table}

None of this is a complaint about the artifact, which ran and produced results
consistent with its paper. It is an observation about what artifact evaluation
measures. Executability gets checked in the environment the authors specify.
Portability across hardware generations, driver stacks and library versions is a
different property, and it is the one that decides whether the work can be built on a
year later by someone with different hardware. About $45\%$ of our code exists only to
construct an environment where the other $55\%$ can run.

\subsection{Detector interfaces decide what can be asked}

The result we most want other researchers to pick up concerns the detector's return
type rather than the attack. Returning a thresholded decision or a CDF value is enough
for deployment and not enough for analysis. Our whole instrumentation effort exists to
undo one line of the released detector: the decision to return $p$ instead of the
intermediate detector quantities $(\xstat, \lam, \mathrm{df}, \sigma)$ that $p$ is
computed from.

Exposing those quantities costs nothing, since they are already computed. Not exposing
them costs a downstream consumer a reimplementation that then has to be validated
against the original, which is exactly the work we did in
Section~\ref{sec:instr}. Detectors meant for research use should return these
intermediate quantities alongside whatever decision they make. This is an argument about
interface design, not about the correctness of the scheme.

\subsection{Predictions that did not survive measurement}

Two of our inferences came from reading code rather than running it, and both were
wrong in the same way.

The first was that the detector's $p$-value is left-censored at zero. The reasoning
was not silly: a non-central $\chi^2$ CDF evaluated deep into its lower tail is a
plausible underflow candidate. Section~\ref{sec:g3} reports the measurement that
disproved it.

The second concerned half precision. The targeted autoencoder upcast resolved the
observed failure, but our controlled probe was run after the patch was already in the
fork (Section~\ref{sec:fp16}). It therefore validates the patched path rather than
independently localising the original fault. A probe that cannot fail is not a test,
and we report the result as the former rather than the latter.

There is a third, smaller correction: we first attributed a PSNR calculation fix to
commit \code{43eec7e}, which touches documentation only. The fix is \code{d5885a3}.

What links them is that in each case we reasoned from what the code appeared to do
rather than from what it measurably did, and in the second case built the check only
after the fix. We report them because that shortcut is available at every point in a
reproduction, and because a study that only surfaces the predictions that held has
not shown it could catch one that did not. Arp et
al.~\cite{arp2022dosdonts} catalogue related failure modes in machine-learning
security research; ours is a variant of their point that assumptions carried into an
evaluation tend to survive it unexamined.

\subsection{What this says about the attack}

Our reproduction supports the original conclusion. Forgery worked in five of six
trials on the first configuration we tried, at \reptime\ per attempt on free
hardware, using a proxy that shares no weights with the target. The attacker needs one
watermarked reference image, and the median forged $p$-value lands three orders of
magnitude below the calibrated threshold.

The property being exploited is specific rather than general. Tree-Ring detection is a
similarity test against a recoverable artifact, and the secret used to verify is the
same secret used to inject. These two properties help explain why this attack shape applies to Tree-Ring and to
related inversion-based semantic watermarking schemes, and the pattern's
transferability across models means the attacker never needs the target model itself.
We reproduced only Tree-Ring, so we make no claim about schemes we did not test, and
this is not a general law relating robustness to forgeability.
Schemes that verify with a secret distinct from the one they inject with are not
implicated, which is precisely the direction MetaSeal~\cite{zhou2026metaseal} takes
with content-dependent cryptographic signatures, and which Gunn et
al.~\cite{gunn2025undetectable} approach differently by selecting initial latents
with a pseudorandom error-correcting code so that watermarked and unwatermarked
outputs are computationally indistinguishable.
What it does mean is that robustness evaluations of this family measure a property
orthogonal to the one a provenance claim rests on.

\section{Limitations}
\label{sec:limits}

We state these plainly, because several of them materially constrain what our numbers
support.

\textbf{A single watermark key.} Every observation shares one Tree-Ring seed
($\code{w\_seed} = 999999$). The variation we report is within-key. These are not
independent draws over the key space, and the separation in Section~\ref{sec:eval}
must not be read as a distribution over watermarks. This is the study's most serious
limitation and the first thing we correct in subsequent work. It matters more here
than it would for a robustness study, since Ci et al.~\cite{ci2024ringid} showed the
multi-key setting is exactly where Tree-Ring's behaviour departs from the single-key
case.

\textbf{No forgery-detection baseline.} We measure what the Tree-Ring detector sees
and do not attempt to separate forged from genuine by other means. Lee et
al.~\cite{lee2026geometric} propose a scheme-agnostic pre-verification detector built
on exactly the geometric deviation our intermediate band is consistent with;
evaluating it on our arms is the natural next comparison and we have not run it.

\textbf{Sample size.} Eighteen observations, six per arm. Enough to establish that the
attack reproduces and the statistic is recoverable; not enough for a distributional
claim. The AUC values should be read as ordering indicators rather than calibrated
estimates, and the gap between our $5/6$ forged detection rate and the reported $0.97$
cannot be resolved at this size.

\textbf{Attack scope.} Reprompt only. The optimisation-based Imprint variant costs
roughly two orders of magnitude more per run by the original authors' own reporting,
which puts a replication with meaningful sample size outside our compute allocation.
We make no claim about how Imprint behaves on this hardware.

\textbf{Precision divergence.} The original results are fp32 throughout; ours are mixed
precision with a patched autoencoder path. Our probe found fp16 and fp32 detection
identical on one image after the patch (Section~\ref{sec:fp16}). That is a narrow
result, and it does not establish that fp16 is numerically neutral for the detection
statistic in general.

\textbf{Excluded target models.} FLUX.1-dev needs \code{bfloat16}, unavailable on the
T4. PixArt-$\Sigma$ is supported by the released code but runs a DPM scheduler rather
than DDIM, and the original work's own ablation shows the scheduler substantially
affects latent recovery quality, so an SDXL/PixArt comparison would confound
architecture with scheduler. We report neither.

\textbf{Pristine images only.} Every observation is of an unprocessed image. Realistic
handling such as recompression, resizing or screenshotting is untested, and both
populations would need to pass through matched channels before anything could be
concluded about behaviour under it.

\textbf{Attacker model substitution.} Stability withdrew the original SD~2.1
repository during this work and we used a mirror. We verified the weights load and
produce the expected attack behaviour, but did not compare them bitwise against the
original release.

\section{Conclusion}

We reproduced the Reprompt semantic watermark forgery attack against Tree-Ring on
Stable Diffusion~XL, using the authors' released implementation on free-tier dual T4 GPUs with substantially less
per-GPU memory than the original evaluation environment. The attack
reproduces: genuine images detected six times out of six, clean images none, forged
images five out of six, at \reptime\ per attempt. The clean arm tracks the original
work closely at a mean $p$ of $0.480$ against $0.472$, which is the arm that most
directly validates the detector as configured.

The finding with the widest reach has to do with the detector's interface
rather than the attack. The released detector computes a non-central $\chi^2$
statistic and returns only its CDF. Recovering that statistic, exactly rather than
approximately, shows that two natural scores built from it separate the forged arm at
AUC $0.861$ and $0.972$ on identical samples, which makes the choice of score
something that has to be fixed before separation is observed. It also shows the forged
arm occupying an intermediate band between the clean null and the genuine arm.
The detector computes that information and then discards it at the moment it returns a
decision.

We also report predictions that failed. We had inferred the detector's $p$-value was
left-censored at zero; across eighteen observations it was never zero, with a minimum
of $5.45 \times 10^{-49}$. We had assumed the half-precision fault localised to the
autoencoder before building anything capable of testing it, and the probe we did build
ran against an already-patched fork. In both cases we trusted what the code appeared
to do over what it measurably did. A reproduction that surfaces only its
confirmed predictions has not shown it could catch one that failed.

The immediate next step is to vary the watermark key across clusters, which is the
binding limitation here, and to pass both genuine and forged populations through
matched laundering channels so the distributions being compared have had identical
handling.

\section*{Author Contributions}

Saifur Rahman Tamim designed the gate protocol, built the environment reconstruction
and provenance pinning, wrote the mixed-precision patch and probe, and implemented the
statistic recovery and its validation. Md Taslimul Hasan Toufique carried out the
post-hoc statistical analysis of the gate data (bootstrap confidence intervals,
pairwise Mann-Whitney tests and effect sizes reported in Section~\ref{sec:eval}),
produced the distributional figure (Fig.~\ref{fig:dist}) and the script that
regenerates it, drafted the related-work section, and ran the notebook sessions
alongside the first author. A.M. Tayeful Islam
advised on experimental design and reviewed the manuscript. All authors contributed
to the analysis and the writing.

\section*{Use of Generative AI}

Generative AI language tools were used during drafting and editing of this manuscript.
All technical claims, measurements, calculations and citations were verified by the
authors against the released artifacts, and the authors take full responsibility for
the content.

\section*{Availability}

Two repositories accompany this work. The reproduction notebook, the gate artifacts
(\code{gate1.json}, \code{gate3.json}, \code{gate4.json},
\code{precision\_probe.json}, \code{phase0\_scores.csv},
\code{phase0\_verdict.json}, \code{session\_meta.json}), the statistical re-analysis
script (\code{stat\_analysis.py}), the figure-generating script and the source of
this paper are at\\
\url{https://github.com/sr-tamim/watermark-forgery-research},
tagged \code{arxiv-v1} at the state described here. The notebook was re-executed in an
independent session on 8 September 2026 from a clean environment build;
\code{session\_meta.json} for both sessions records the same fork commit and the same
dependency-lock SHA-256.

The modified attack code is a fork of M\"uller et al.'s released
repository~\cite{semanticforgery-repo}, at\\
\url{https://github.com/sr-tamim/semantic-forgery}, pinned for all runs reported here
at commit \code{7f9e7ad}, which branches from upstream \code{ca68950}. Both hashes
are recorded in \code{session\_meta.json} alongside the SHA-256 of the resolved
dependency lock, so any run can be traced to the exact code that produced it.

\balance
\bibliographystyle{plainnat}
\bibliography{refs}

@inproceedings{muller2025semantic,
  title     = {Black-Box Forgery Attacks on Semantic Watermarks for
               Diffusion Models},
  author    = {M{\"u}ller, Andreas and Lukovnikov, Denis and Thietke, Jonas
               and Fischer, Asja and Quiring, Erwin},
  booktitle = {IEEE/CVF Conference on Computer Vision and Pattern Recognition
               (CVPR)},
  year      = {2025},
  note      = {arXiv:2412.03283},
}

@inproceedings{wen2023treering,
  title     = {Tree-Rings Watermarks: Invisible Fingerprints for Diffusion
               Images},
  author    = {Wen, Yuxin and Kirchenbauer, John and Geiping, Jonas and
               Goldstein, Tom},
  booktitle = {Advances in Neural Information Processing Systems (NeurIPS)},
  year      = {2023},
}

@inproceedings{song2021ddim,
  title     = {Denoising Diffusion Implicit Models},
  author    = {Song, Jiaming and Meng, Chenlin and Ermon, Stefano},
  booktitle = {International Conference on Learning Representations (ICLR)},
  year      = {2021},
}

@inproceedings{podell2024sdxl,
  title     = {{SDXL}: Improving Latent Diffusion Models for High-Resolution
               Image Synthesis},
  author    = {Podell, Dustin and English, Zion and Lacey, Kyle and
               Blattmann, Andreas and Dockhorn, Tim and M{\"u}ller, Jonas and
               Penna, Joe and Rombach, Robin},
  booktitle = {International Conference on Learning Representations (ICLR)},
  year      = {2024},
}

@inproceedings{rombach2022ldm,
  title     = {High-Resolution Image Synthesis with Latent Diffusion Models},
  author    = {Rombach, Robin and Blattmann, Andreas and Lorenz, Dominik and
               Esser, Patrick and Ommer, Bj{\"o}rn},
  booktitle = {IEEE/CVF Conference on Computer Vision and Pattern Recognition
               (CVPR)},
  year      = {2022},
}

@inproceedings{yang2024gaussianshading,
  title     = {Gaussian Shading: Provable Performance-Lossless Image
               Watermarking for Diffusion Models},
  author    = {Yang, Zijin and Zeng, Kai and Chen, Kejiang and Fang, Han and
               Zhang, Weiming and Yu, Nenghai},
  booktitle = {IEEE/CVF Conference on Computer Vision and Pattern Recognition
               (CVPR)},
  pages     = {12162--12171},
  year      = {2024},
}

@article{zhu2025optfree,
  title   = {Optimization-Free Universal Watermark Forgery with Regenerative
             Diffusion Models},
  author  = {Zhu, Chaoyi and Li, Zaitang and Yang, Renyi and Birke, Robert and
             Chen, Pin-Yu and Ho, Tsung-Yi and Chen, Lydia Y.},
  journal = {arXiv preprint arXiv:2506.06018},
  year    = {2025},
}

@article{zhou2026metaseal,
  title   = {{MetaSeal}: Defending Against Image Attribution Forgery Through
             Content-Dependent Cryptographic Watermarks},
  author  = {Zhou, Tong and Ding, Ruyi and Liu, Gaowen and Fleming, Charles and
             Kompella, Ramana Rao and Fei, Yunsi and Xu, Xiaolin and Ren, Shaolei},
  journal = {Transactions on Machine Learning Research},
  year    = {2026},
  note    = {arXiv:2509.10766},
}

@inproceedings{arp2022dosdonts,
  title     = {Dos and Don'ts of Machine Learning in Computer Security},
  author    = {Arp, Daniel and Quiring, Erwin and Pendlebury, Feargus and
               Warnecke, Alexander and Pierazzi, Fabio and Wressnegger,
               Christian and Cavallaro, Lorenzo and Rieck, Konrad},
  booktitle = {USENIX Security Symposium},
  year      = {2022},
}

@misc{semanticforgery-repo,
  title        = {semantic-forgery},
  author       = {M{\"u}ller, Andreas},
  howpublished = {\url{https://github.com/and-mill/semantic-forgery}},
  note         = {Accessed August 2026; pinned at commit \texttt{ca68950}},
  year         = {2024},
}

@inproceedings{olszewski2023reproducibility,
  title     = {``Get in Researchers; We're Measuring Reproducibility'':
               A Reproducibility Study of Machine Learning Papers in
               Tier 1 Security Conferences},
  author    = {Olszewski, Daniel and Lu, Allison and Stillman, Carson and
               Warren, Kevin and Kitroser, Cole and Pascual, Alejandro and
               Ukyab, Divyajyoti and Traynor, Patrick and Butler, Kevin},
  booktitle = {ACM SIGSAC Conference on Computer and Communications Security
               (CCS)},
  year      = {2023},
}

@inproceedings{an2024waves,
  title     = {Benchmarking the Robustness of Image Watermarks},
  author    = {An, Bang and Ding, Mucong and Rabbani, Tahseen and
               Agrawal, Aakriti and Xu, Yuancheng and Deng, Chenghao and
               Zhu, Sicheng and Mohamed, Abdirisak and Wen, Yuxin and
               Goldstein, Tom and Huang, Furong},
  booktitle = {International Conference on Machine Learning (ICML)},
  year      = {2024},
  note      = {arXiv:2401.08573},
}

@article{hwang2024invisible,
  title   = {Invisible Watermarks: Attacks and Robustness},
  author  = {Hwang, Dongjun and Woo, Sungwon and Gao, Tom and
             Luo, Raymond and Baek, Sunghwan},
  journal = {arXiv preprint arXiv:2412.12511},
  year    = {2024},
}

@inproceedings{zhao2024removable,
  title     = {Invisible Image Watermarks Are Provably Removable Using
               Generative {AI}},
  author    = {Zhao, Xuandong and Zhang, Kexun and Su, Zihao and Vasan, Saastha
               and Grishchenko, Ilya and Kruegel, Christopher and Vigna, Giovanni
               and Wang, Yu-Xiang and Li, Lei},
  booktitle = {Advances in Neural Information Processing Systems (NeurIPS)},
  year      = {2024},
}

@inproceedings{ci2024ringid,
  title     = {{RingID}: Rethinking Tree-Ring Watermarking for Enhanced
               Multi-Key Identification},
  author    = {Ci, Hai and Yang, Pei and Song, Yiren and Shou, Mike Zheng},
  booktitle = {European Conference on Computer Vision (ECCV)},
  pages     = {338--354},
  year      = {2024},
}

@inproceedings{gunn2025undetectable,
  title     = {An Undetectable Watermark for Generative Image Models},
  author    = {Gunn, Sam and Zhao, Xuandong and Song, Dawn},
  booktitle = {International Conference on Learning Representations (ICLR)},
  year      = {2025},
  note      = {arXiv:2410.07369},
}

@inproceedings{fernandez2023stablesignature,
  title     = {The Stable Signature: Rooting Watermarks in Latent Diffusion
               Models},
  author    = {Fernandez, Pierre and Couairon, Guillaume and J{\'e}gou, Herv{\'e}
               and Douze, Matthijs and Furon, Teddy},
  booktitle = {IEEE/CVF International Conference on Computer Vision (ICCV)},
  pages     = {22466--22477},
  year      = {2023},
}

@inproceedings{zhu2018hidden,
  title     = {{HiDDeN}: Hiding Data with Deep Networks},
  author    = {Zhu, Jiren and Kaplan, Russell and Johnson, Justin and
               Fei-Fei, Li},
  booktitle = {European Conference on Computer Vision (ECCV)},
  pages     = {657--672},
  year      = {2018},
}

@inproceedings{ho2020ddpm,
  title     = {Denoising Diffusion Probabilistic Models},
  author    = {Ho, Jonathan and Jain, Ajay and Abbeel, Pieter},
  booktitle = {Advances in Neural Information Processing Systems (NeurIPS)},
  year      = {2020},
}

@book{cox2007digital,
  title     = {Digital Watermarking and Steganography},
  author    = {Cox, Ingemar J. and Miller, Matthew L. and Bloom, Jeffrey A. and
               Fridrich, Jessica and Kalker, Ton},
  publisher = {Morgan Kaufmann},
  edition   = {2nd},
  year      = {2007},
}

@article{wilson1927interval,
  title   = {Probable Inference, the Law of Succession, and Statistical
             Inference},
  author  = {Wilson, Edwin B.},
  journal = {Journal of the American Statistical Association},
  volume  = {22},
  number  = {158},
  pages   = {209--212},
  year    = {1927},
}

@inproceedings{lee2026geometric,
  title        = {Rethinking Forgery Attacks on Semantic Watermarks in Black-Box
                  Settings: A Geometric Distortion Perspective},
  author       = {Lee, Cheng-Yi and Zhang, Yichi and Yang, Yuchen and
                  Lu, Chun-Shien and Chen, Jun-Cheng},
  booktitle    = {International Conference on Machine Learning (ICML)},
  year         = {2026},
  note         = {arXiv:2606.29807},
}

@inproceedings{zhang2026sembind,
  title        = {SemBind: Binding Diffusion Watermarks to Semantics Against
                  Black-Box Forgery Attacks},
  author       = {Zhang, Xin and Yang, Zijin and Chen, Kejiang and Ma, Linfeng and
                  Zhang, Weiming and Yu, Nenghai},
  booktitle    = {International Conference on Machine Learning (ICML)},
  year         = {2026},
  note         = {arXiv:2601.20310},
}

\end{document}